\documentclass[twocolumn,showpacs,preprintnumbers,amsmath,amssymb,prl]{revtex4}

\usepackage{graphicx}% Include figure files
\usepackage{dcolumn}% Align table columns on decimal point
\usepackage{bm}% bold math

\begin{document}

%\preprint{S.~Kasahara {\it et al}.,}

\title{Evolution from Non-Fermi to Fermi Liquid Transport Properties by Isovalent Doping in BaFe$_{2}$(As$_{1-x}$P$_{x}$)$_2$ Superconductors
}

\author{S.~Kasahara$^{1,3,\ast}$}%\email{kasa@scphys.kyoto-u.ac.jp} 
%\homepage{http://www.nims.jp/filmcrystal/kasahara/kasahara.html}
\author{T.~Shibauchi$^{2}$}
\author{K.~Hashimoto$^{2}$}
\author{K.~Ikada$^{2}$}
\author{S.~Tonegawa$^{2}$}
\author{R.~Okazaki$^{2}$}
\author{H.~Ikeda$^{2}$}
\author{H.~Takeya$^{3}$}
\author{K.~Hirata$^{3}$}
\author{T.~Terashima$^{1}$}
\author{Y.~Matsuda$^{2}$}

\affiliation{$^{1}$
Research Center for Low Temperature and Materials Sciences, Kyoto University, Kyoto 606-8502, Japan
}

\affiliation{$^{2}$
Department of Physics, Kyoto University, Kyoto 606-8502, Japan
}

\affiliation{$^{3}$
Superconducting Materials Center, National Institute for Materials Science, Tsukuba, Ibaraki 305-0047, Japan
}

\date{\today}% It is always \today, today,

\begin{abstract}
The normal-state charge transport is studied systematically in high-quality single crystals of BaFe$_2$(As$_{1-x}$P$_x$)$_2$ ($0 \leq x \leq 0.71$). By substituting isovalent P for As, the spin-density-wave (SDW) state is suppressed and the dome-shaped superconducting phase ($T_c \lesssim 31$\,K) appears. Near the SDW end point ($x\approx0.3$), we observe striking linear temperature ($T$) dependence of resistivity in a wide $T$-range, and remarkable low-$T$ enhancement of Hall coefficient magnitude from the carrier number estimates.  We also find that the magnetoresistance apparently violates the Kohler's rule and is well scaled by the Hall angle $\Theta_H$ as $\Delta\rho_{xx}/\rho_{xx} \propto \tan^2\Theta_H$. These non-Fermi liquid transport anomalies cannot be attributed to the simple multiband effects. These results capture universal features of correlated electron systems in the presence of strong antiferromagnetic fluctuations.

\end{abstract}

\pacs{
74.70.Dd %Ternary, quaternary, and multinary compounds (including Chevrel phases, borocarbides, etc.)
74.25.Fy %Transport properties
74.25.Dw %Superconductivity phase diagrams 
74.25.Jb %Electronic structure 
%74.62.-c %Transition temperature variations 
%74.62.Bf %Effects of material synthesis, crystal structure, and chemical composition 
%74.62.Dh %Effects of crystal defects, doping and substitution 
%74.62.Fj %Pressure effects 
%74.62.Yb %Other effects
}  
%74.25.Bt Thermodynamic properties
\maketitle

Within the last decade, it has been found that in strongly correlated electron systems various transport properties display striking deviations from the conventional Fermi-liquid behavior, particularly in the vicinity of magnetic instability.  It is generally believed that strong magnetic fluctuations seriously modify the quasiparticle masses and scattering cross section of the Fermi-liquid.  Very recently discovered Fe-pnictides \cite{Kamihara_JACS} have stimulated great interest because the high-$T_c$ superconductivity with nontrivial Cooper pairing state occurs in the vicinity of the spin-density-wave (SDW) instability. It has been suggested that the antiferromagnetic fluctuations associated with the Fermi surface nesting between electron- and hole-pockets are essential for the occurrence of the superconductivity \cite{Mazin, Kuroki, Ikeda}. Therefore clarifying the normal state electron transport of Fe-pnictides is of crucial importance, since it might be a key to elucidating the mechanism of high-$T_c$ superconductivity.

Until now, systematic transport studies of Fe-pnictides have been reported mainly for polycrystalline samples such as SmFeAs(O,F) \cite{Liu08} and (K,Sr)Fe$_2$As$_2$ \cite{Gooch09}. More recently, single crystals of Ba(Fe,Co)$_2$As$_2$ become available and the transport properties show anomalous behaviors \cite{Wang09,Fang09,Rullier09,Doiron09}.  However, the interpretations are yet controversial as to the importance of magnetic fluctuations and multiband effects.  This may arise from the multiple effects of charge doping, which makes imbalance between holes and electrons.  Moreover, the scattering mechanism of Co atoms substituted within the Fe-planes is unclear.

Here we present a systematic study of the dc-resistivity, Hall effect and magnetoresistance in the normal state of high-quality single crystals of BaFe$_2$(As$_{1-x}$P$_x$)$_2$, ranging from the undoped SDW ($x=0$) state to the overdoped Fermi-liquid state ($x\approx0.71$) through the highly unusual non-Fermi liquid state. By substituting isovalent P for As, the SDW state is suppressed and the superconducting phase ($T_c \lesssim 31$\,K) appears \cite{Ba-P}.  BaFe$_2$(As$_{1-x}$P$_x$)$_2$ is the most suitable system to study the transport properties of Fe-pnictides, because of the following reasons.  Firstly, the system can be assumed as a compensated metal, i.e. essentially the same number of electrons and holes, $n_e=n_h=n$ for any $x$-value, which is supported by the band structure calculations \cite{compensation}.  Therefore we can only tune magnetic character without nominally changing charge carrier concentrations.  This makes the interpretation of transport coefficients much simpler than the case for charge doping.  Indeed, it is theoretically suggested that P doping of As presents a means to access a novel type of magnetic quantum criticality in an unmasked fashion \cite{Dai}.  Secondly, we succeeded to grow very clean single crystals with large residual resistivity ratio values ($\approx 25$ for $x = 0.41$). In fact, in recent high-field studies clear quantum oscillations are observed in our doped crystals at $x\agt0.4$ \cite{Shishido_dHvA}, indicating that the substitution in the pnictogen sites induces less impurity scattering than that in the Fe-planes \cite{Wang09,Fang09,Rullier09,Doiron09}. By using these clean crystals, we observe that all of the transport coefficients exhibit striking deviations from the conventional Fermi-liquid behaviors near the SDW end point. These anomalies become less pronounced as the nesting conditions degrade with overdoping. We show that these non-Fermi liquid transport anomalies cannot be attributed to the simple multiband effects. Several noticeable features highlight common non-Fermi-liquid behaviors in strongly correlated electron systems in the vicinity of antiferromagnetism.

\begin{figure}[t]
%h=here, t=top, b=bottom, p=separate figure pag
\begin{center}\leavevmode
\includegraphics[width=85mm]{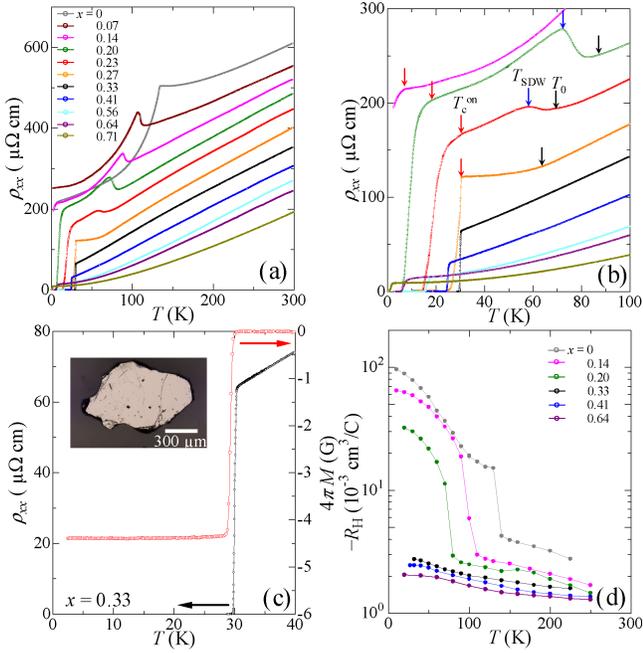}
\caption{
(Color online)
(a) $\rho_{xx}(T)$ curves of BaFe$_2$(As$_{1-x}$P$_{x}$)$_{2}$ in zero field. (b) The same data below 100\,K. 
(c) $\rho_{xx}(T)$ and magnetization $M(T)$ measured by SQUID magnetometer near $T_c$ for $x=0.33$. The inset is a photograph of the (001) surface of a single crystal. 
(d) $T$-dependence of the Hall coefficient $R_H(T)$ for various doping levels. 
}
\label{RT}
\end{center}
\end{figure}

Single crystals of BaFe$_2$(As$_{1-x}$P$_x$)$_2$ were grown from stoichiometric mixtures of Ba (flakes), and FeAs, Fe, P, or FeP (powders) placed in an alumina crucible, sealed in an evacuated quartz tube. It was heated up to 1150--1200$^\circ$C, kept for 12 hours, and then cooled slowly down to $\sim 800^\circ$C at the rate of 1.5$^\circ$C/hours. Platelet crystals with shiny (001) surface were extracted [inset of Fig.\:\ref{RT}(c)]. $x$-values were determined by an energy dispersive X-ray analyzer. Band structure was calculated by density functional theory implemented in the WIEN2k code \cite{WIEN2k}.

%\section{Results and discussion}

%%%%%%%%%%%%%%%%%%%%%%FIG 1%%%%%%%%%%%%%%%
%%%%%%%%%%%%%%%%%%%%%%FIG 1%%%%%%%%%%%%%%%

Figures\:\ref{RT}(a) and (b) show the in-plane resistivity $\rho_{xx}(T)$ in zero field below 300\,K and 100\,K, respectively. An anomaly in $\rho_{xx}(T)$ in the parent BaFe$_{2}$As$_2$ at $T_0 =137$\,K corresponds to the structural and simultaneous SDW transitions ($T_0=T_{\rm SDW}$), consistent with the previous studies \cite{Rotter_BaFe2As2, Wang_BaFe2As2_single}. With increasing $x$, the anomaly is replaced by a step like increase at $T_0$, followed by a sharp peak at $T_{\rm SDW}$. The increase of $x$ suppresses these anomalies towards lower temperatures.  At the same time, the resistivity shows a drop at lower temperatures and zero resistivity is attained at $x\geq 0.20$, indicating a coexistence of SDW and superconductivity. The coexistence is observed up to $x=0.28$, and no anomaly associated with the SDW transition is observed for $x\geq 0.33$. As shown in Fig.\:\ref{RT}(c), the crystals exhibit a very narrow superconducting transition width ($\Delta T_c<0.4$\,K for $x=0.33$).

%%%%%%%%%%%%%%%%%%%%%%FIG 2%%%%%%%%%%%%%%%
\begin{figure}[t]
%h=here, t=top, b=bottom, p=separate figure page
\begin{center}\leavevmode
\includegraphics[width=75mm]{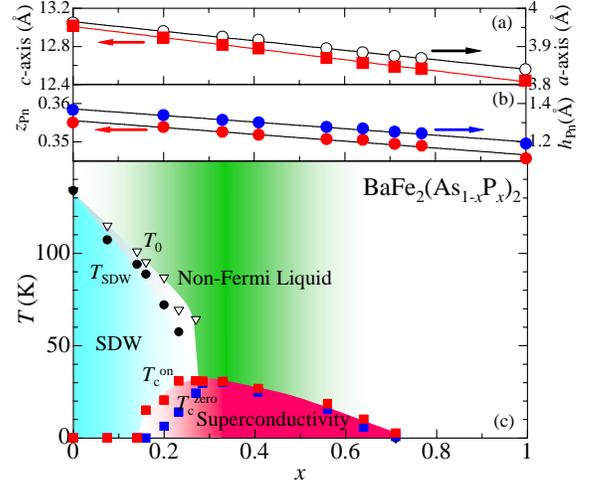}
\caption{
(Color online)
(a) Lattice constants determined from X-ray as a function of $x$. (b) The $z$ coordinate of pnictogen atoms in the unit cell $z_{\rm Pn}$ and the pnictogen height from the iron plane $h_{\rm Pn}=(z_{\rm Pn}-0.25)\times c$. (c) Phase diagram of BaFe$_2$(As$_{1-x}$P$_{x}$)$_{2}$ against the P content $x$. The open triangles show the structural transition at $T_0$. The closed black circles show $T_{\rm SDW}$, where the resistivity show reductions due to the reduced spin scattering at the SDW transition. The onset of superconductivity, $T_c^{\rm on}$, and the zero resistivity temperature, $T_c^{\rm zero}$, are displayed as blue and red squares.
 }
\label{TvsX}
\end{center}
\end{figure}
%%%%%%%%%%%%%%%%%%%%%%FIG 2%%%%%%%%%%%%%%%

Figure\:\ref{RT}(d) depicts the temperature dependence of the Hall coefficient $R_H$ at various doping levels.  $R_H$ is defined as the field derivative of Hall resistivity $\rho_{xy}$, $R_H \equiv d\rho_{xy}/dH$, at $\mu_0H \rightarrow0$\,T. $R_H$ is negative, indicating the dominant contribution of electrons. For $x\leq 0.14$, with decreasing temperature $|R_H|$ jumps to higher values at $T_{\rm SDW}$, indicating a reduction of carriers due to the development of SDW gap in the Fermi surface.  $R_H$ is strongly temperature dependent even in the non-magnetic phase.

Figure\:\ref{TvsX}(c) displays the $T$-$x$ phase diagram of the BaFe$_2$(As$_{1-x}$P$_{x}$)$_2$ system obtained in the present study. $T_{\rm SDW}$ decreases rapidly with $x$. At $0.14\alt x \alt 0.30$, both the SDW and superconducting transitions are observed, suggesting that the superconductivity may not have a bulk character. In the vicinity of $x\approx 0.30$, the SDW transition disappears suddenly and a rapid growth of bulk superconductivity appears. The superconducting transition shows a maximum $T_c=31$\,K at $x=0.26$ and further increase of $x$ leads to the reduction of $T_c$. The obtained phase diagram bears striking resemblance to that of the pressure dependence of BaFe$_2$As$_2$ \cite{Ba_122_pressure}, indicating that the isovalent substitution of P for As is identical to the pressure effect. In fact, the lattice constants as well as the pnictogen height decrease linearly with $x$, as shown in Figs.\:\ref{TvsX}(a) and (b).

%%%%%%%%%%%%%%%%%%%%%%FIG 3%%%%%%%%%%%%%%%
\begin{figure}[t]
%h=here, t=top, b=bottom, p=separate figure page
\begin{center}\leavevmode
\includegraphics[width=85mm]{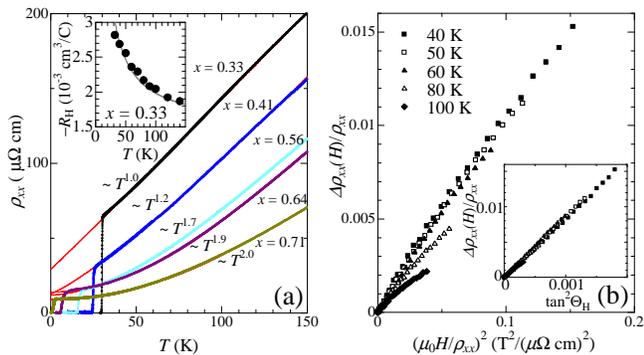}

\caption{
(Color online)
(a) Normal-state $\rho_{xx}(T)$ for $x=0.33$, 0.41, 0.56, 0.64, and 0.71 at low temperatures can be fitted by the power-law (Eq.\:(\ref{rho})). The inset shows $T$-dependence of $-R_H(T)$ for $x=0.33$. Solid line is a fit to the data by Eq.\:\ref{Halleq} with $C_1=0.048$\,Kcm$^3$/C and $C_2=1.5 \times 10^{-3}$\,cm$^{3}$/C. 
(b) Magnetoresistance $\Delta\rho_{xx}(H)/\rho_{xx}$ plotted as a function of $(\mu_0H/\rho_{xx})^2$ for $x=0.33$.  The inset shows $\Delta\rho_{xx}(H)/\rho_{xx}$ plotted as a function $\tan^2\Theta_H$. 
}
\label{nFL}
\end{center}
\end{figure}
%%%%%%%%%%%%%%%%%%%%%%FIG 3%%%%%%%%%%%%%%%

Now we discuss the normal-state transport properties focusing on the non-magnetic regime.  At $x=0.33$ just beyond the SDW end point, $\rho_{xx}(T)$ exhibits a nearly perfect $T$-linear dependence in a wide $T$-range above $T_c$ as shown in Fig.\:\ref{nFL}(a); 
\begin{equation}
\rho_{xx}(T) = \rho_0+AT^{\alpha} 
\label{rho}
\end{equation}
with $\alpha=1.0$, where $A$ is a constant \cite{Ioffe}. Thus the resistivity exhibits a striking deviation from the standard Fermi-liquid theory with $\alpha=2$.  With increasing $x$, $\alpha$ increases and the Fermi-liquid behavior is recovered at $x=0.71$. For $x=0.33$, $R_H$ exhibits a marked $T$-dependence that is approximated as  
\begin{equation}
-R_{H}(T) = C_1/T+C_2, 
\label{Halleq}
\end{equation}
where $C_1$ and $C_2$ are positive constants, as depicted in the inset of Fig.\:\ref{nFL}(a). Similar anomalous behaviors of $\rho_{xx}$ and $R_H$ are reported in other Fe-pnictides \cite{Liu08,Gooch09,Wang09,Fang09,Rullier09}. It is well known that temperature-dependent $R_H$ can be obtained in the Bloch theory when multiple bands are involved. Then an important question is whether the most fundamental transport properties described by Eqs.\:(\ref{rho}) and (\ref{Halleq}) can be accounted for by the conventional multiband model or can be indicative of novel transport properties inherent to the Fe-based systems. We show that the former is highly unlikely for the following reasons.

We note that the compensation condition in the present isovalent system is important because it allows us to simply describe 
%the diagonal conductivity $\sigma (\approx 1/\rho_{xx})$ and 
the Hall coefficient as
%\begin{eqnarray}
%\sigma &=& \sigma_h + \sigma_e, \\
%R_H &=& \frac{1}{ne} \times \frac{\sigma_h - \sigma_e}{\sigma_h + \sigma_e}, 
 \label{twoband}
%\end{eqnarray}
\begin{equation}
R_H = \frac{1}{ne} \times \frac{\sigma_h - \sigma_e}{\sigma_h + \sigma_e}, 
 \label{twoband}
\end{equation}
where $\sigma_e$ ($\sigma_h$) is the conductivity of electron (hole) band.
The fact that $R_H$ is negative indicates that the electron band dominates transport properties ($\sigma_e>\sigma_h$). Strong evidence against the multiband explanation is obtained from the amplitude of $R_H$. From Eq.\:(\ref{twoband}), $|R_H|$ cannot exceed $1/ne$. Band calculations reveal that BaFe$_2$As$_2$ has $\sim 0.15$ electrons per Fe \cite{Fang09}, and that the electron Fermi surface is not seriously influenced by the P replacement (see below). This electron density corresponds to $1/ne\approx 0.98\times 10^{-3}$\,cm$^3$/C. However, it is clear from the inset of Fig.\:\ref{nFL}(a) that the observed magnitude of $R_{\rm H}$ becomes considerably larger than this value especially at low temperatures. These results lead us to conclude that the simple multiband picture cannot explain the transport coefficients in the present system. 

Another anomalous feature is also found in magnetoresistance (MR). In the Fermi liquid state, the MR, $\Delta \rho_{xx}(H)/\rho_{xx}\equiv (\rho_{xx}(H)-\rho_{xx}(H=0))/\rho_{xx}$, obeys Kohler's rule, $\Delta \rho_{xx}(H)/\rho_{xx}=F(\mu_0 H/\rho_{xx})$, where $F(y)$ is a function of $y$ depending on the electronic structure.  Figure\:\ref{nFL}(b) is the transverse MR plotted against $\mu_0 H/\rho_{xx}$ for $x=0.33$ in {\boldmath $H$} $\parallel c$. The data at different temperatures are on distinctly different curves, indicating apparent violation of the Kohler's rule.   It has been proposed \cite{Kontani,Varma} that the MR in the non-Fermi-liquid regime may be scaled by the  Hall angle $\Theta_H(\equiv \tan^{-1} \frac{\rho_{xy}}{\rho_{xx}})$ as $\Delta\rho_{xx}(H)/\rho_{xx} \propto \tan^2\Theta_H$ (modified Kohler's rule).  To examine this relation,  we plot the MR as a function of $\tan^2 \Theta_H$ in the inset of Fig.\:\ref{nFL}(b).  Obviously, the MR data at different temperatures collapse into the same curve, indicating a distinct Hall angle scaling of the MR.

It should be noted that the modified Kohler's rule as well as the $T$-linear $\rho_{xx}$ and the low-temperature enhancement of $|R_{H}|(\gg1/ne)$, distinct from the standard Fermi-liquid theory of metals, have also been reported in other strongly correlated electron systems including high-$T_c$ cuprates \cite{PP-HTSC-II} and 2D heavy fermion compounds \cite{Nakajima07}. The simultaneous understanding of these anomalies has been a subject of intense research \cite{Stojkovic97, Varma, Kontani, Prelovsek}. Among others, one may involve different quasiparticle scattering times $\tau$ at different parts of Fermi surface \cite{Stojkovic97, Varma}. The effects of band curvature and Fermi velocity anisotropy on $\tau$ can account for the enhancement of $|R_{\rm H}|$. Another important effect is the vertex corrections to the longitudinal and transverse conductivities due to large antiferromagnetic fluctuations \cite{Kontani}, which modify the current at the Fermi surface spots connecting with the nesting vectors \cite{Nakajima07}. Although quantitative analysis of these effects in the Fe-based superconductors deserves further theoretical studies, the Hall-angle scaling of the MR can be consistently explained by these theories \cite{Varma,Kontani}. In any case, the observed non-Fermi-liquid properties strongly suggest the significance of antiferromagnetic fluctuations in the transport coefficients.   When we go away from the antiferromagnetism by overdoping, these odd properties indeed become less pronounced as shown in Fig.\:\ref{nFL}(a), consistent with the above view. 

%%%%%%%%%%%%%%%%%%%%%%FIG 4%%%%%%%%%%%%%%%
\begin{figure}[t]
%h=here, t=top, b=bottom, p=separate figure page
\begin{center}\leavevmode
\includegraphics[width=85mm]{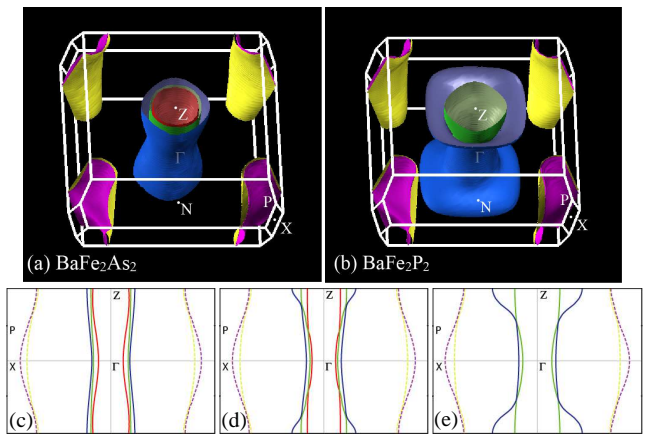}
\caption{
(Color online)
Fermi surfaces of (a) BaFe$_2$As$_2$ ($x = 0$) and (b) BaFe$_2$P$_2$ ($x = 1$).  Cross-section views along (110) plane are also shown for (c) $x = 0$, (d) 0.3, and (e) 1. For $x=0.3$, we use the BaFe$_2$As$_2$ structure with the experimental atomic positions. 
}
\label{FS}
\end{center}
\end{figure}
%%%%%%%%%%%%%%%%%%%%%%FIG 4%%%%%%%%%%%%%%%

The effect of antiferromagnetic fluctuations on the transport properties is also suggested by electronic structure calculations. Figure\:\ref{FS} displays Fermi surfaces calculated for BaFe$_2$(As$_{1-x}$P$_x$)$_2$ ($x = 0$, 0.3, and 1). Three hole sheets exist around $\Gamma$ in BaFe$_2$As$_2$, while one of them is absent in BaFe$_2$P$_2$. Both compounds have two electron pockets around $X$. Three dimensionality of the hole Fermi surfaces is quite sensitive to the pnictogen position $z_{\rm Pn}$. Similar sensitivity has also been pointed out in LaFePO \cite{Vildosola_LaFeAsO_LaFePO} and SrFe$_2$P$_2$ \cite{Analytis_SrFe2P2}. In the present BaFe$_2$(As$_{1-x}$P$_{x}$)$_2$ system, the substitution of P for As is expected to induce a small reduction of $z_{\rm Pn}$ [Fig.\:\ref{TvsX}(b)], which dramatically promotes the three dimensionality of the hole sheets. The enlarged Fermi surface warping upon doping is considered to weaken the nesting along the $(\pi, \pi)$ direction, minify the SDW phase, and then induce the superconductivity. In contrast to the significant change in the hole sheets, the electron sheets, which are the dominant carriers for the transport, are almost unchanged. Thus the large change in the antiferromagnetic nesting conditions is the most relevant to the evolution from the non-Fermi to Fermi-liquid properties by the isovalent doping in this system.

In summary, we have systematically studied the electronic transport properties of BaFe$_2$(As$_{1-x}$P$_{x}$)$_2$ using high-quality single crystals, ranging from non-doped SDW to overdoped Fermi-liquid states. Near the SDW end point, anomalous non-Fermi-liquid behaviors, including a pronounced $T$-linear $\rho_{xx}$, a striking enhancement of Hall coefficient at low $T$ ($|R_H|\gg 1/ne$), and violation of the Kohler's rule and Hall-angle scaling in the magnetoresistance, are observed. These cannot be attributed to the simple multiband effect. The doping dependence points to an important role of antiferromagnetic fluctuations in the non-Fermi-liquid transport properties in the Fe-pnictides. 
These highly unusual transport properties, commonly observed in Fe-pnictides, high-$T_c$ cuprates, and heavy-fermion superconductors, very likely capture universal features of correlated electron system near the magnetic instability.

%\begin{acknowledgments}
We thank R. Hashimoto, H. Hayashi, I. Tanaka, A. Kitada, H. Yamochi for technical help, and R. Arita, K. Ishida, H. Kontani, D. Pines, J. Schmalian and T. Tohyama for valuable discussions. 
This work is partially supported by KAKENHI and Grant-in-Aid for GCOE program 
``The Next Generation of Physics, Spun from Universality and Emergence'' from MEXT, Japan.


\begin{thebibliography}{99}
\bibitem{compensation} 
BaFe$_2$As$_2$ has even number (174) of total electrons per unit formula, and the substitution of an As atom with the congener P only changes the number of electrons in the closed shells with keeping the number of conduction electrons the same. The band calculations therefore result in the compensated electronic structure.
\bibitem{Ioffe}
The Ioffe-Regel limit is roughly estimated as $\rho\sim 350$\,$\mu\Omega$cm, which may be relevant to the deviation from the $T$-linear behavior of $\rho(T)$ at high $T$ ($>150$\,K).
\end{thebibliography}
\end{document}